\documentclass[prl,amsmath,amssymb,twocolumn]{revtex4-2}

\pdfoutput=1

\usepackage{graphicx}
\usepackage{subfigure}
\usepackage{adjustbox}
\usepackage{bm}
\usepackage{color}
\usepackage{braket}
\usepackage{standalone}
\usepackage{multirow}
\usepackage{tikz}
\usepackage{mathrsfs}
\usepackage{dsfont}
\usepackage[colorlinks,bookmarks=true,citecolor=blue,linkcolor=red,urlcolor=blue]{hyperref}

\newcommand{\bea}{\begin{eqnarray}}
\newcommand{\eea}{\end{eqnarray}}
\usepackage{float}
\usepackage{array}
\usepackage{slashed,bbold}

\usepackage{amsmath,amssymb,bm,bbm} 
\usepackage{graphicx}

\usepackage[papersize={8.5in,11in}]{geometry}

\definecolor{darkblue}{rgb}{0.,0.,0.4}
\definecolor{darkred}{rgb}{0.5,0.,0.}
\definecolor{BlueViolet}{RGB}{138,43,226}
\definecolor{SkyBlue}{RGB}{30,144,255}
\definecolor{DarkGreen}{RGB}{0,100,0}
\def \nn{\nonumber \\}

\begin{document}

\title{Symmetry and Higher-Order Exceptional Points}

\author{Ipsita Mandal$^{1,2}$ and Emil J. Bergholtz$^{2}$}

\affiliation{$^1$Institute of Nuclear Physics, Polish Academy of Sciences, 31-342 Krak\'{o}w, Poland\\
$^2$Department of Physics, Stockholm University, AlbaNova University Center, 106 91 Stockholm, Sweden}

\begin{abstract}
{Exceptional points (EPs), at which both eigenvalues and eigenvectors coalesce, are ubiquitous and unique features of non-Hermitian systems. Second-order EPs are by far the most studied due to their abundance, requiring only the tuning of two real parameters, which is less than the three parameters needed to generically find ordinary Hermitian eigenvalue degeneracies. Higher-order EPs generically require more fine-tuning, and are thus assumed to play a much less prominent role. Here, however, we illuminate how physically relevant symmetries make higher-order EPs dramatically more abundant and conceptually richer. More saliently, third-order EPs generically require only two real tuning parameters in the presence of either a parity-time (PT) symmetry or a generalized chiral symmetry. Remarkably, we find that these different symmetries yield topologically distinct types of EPs. We illustrate our findings in simple models, and show how third-order EPs with a generic $\sim k^{1/3}$ dispersion are protected by PT symmetry, while third-order EPs with a $\sim k^{1/2}$ dispersion are protected by the chiral symmetry emerging in non-Hermitian Lieb lattice models. More generally, we identify stable, weak, and fragile aspects of symmetry-protected higher-order EPs, and tease out their concomitant phenomenology.}
\end{abstract}

\maketitle

{\it Introduction.---} With the advent of non-Hermitian (NH) topological phases \cite{review}, exceptional points (EPs) have become an interdisciplinary frontier of research, with applications ranging from classical metamaterials to quantum condensed matter systems \cite{EPreview,review}. Representing the natural dissipative counterpart of stable nodal points \cite{BerryDeg,Heiss} familiar from Weyl semimetals in the Hermitian realm, EPs exhibit fascinating topologically stable phenomena, including the formation of bulk Fermi arcs \cite{NHarc} and unidirectional lasing \cite{lasing}. So far, most work has focused on the simplest example of second-order EPs, i.e. two-fold degenerate eigenvalues sharing a single eigenstate, which naturally occur in two-dimensional (2D) NH systems \cite{koziifu,NHarc}, and are promoted to closed exceptional lines in 3D \cite{EPrings,CaBe2018,EPringExp}. More generally, for obtaining an $n$-th order EP, $2n-2$ real parameters, such as components of a lattice momentum, need to be adjusted (co-dimension $2n-2$). This suggests that $n \ge 3$ cannot be expected to occur in physical systems with up to three spatial dimensions. However, for the aforementioned case of $n=2$, generic NH symmetries have been found to reduce the co-dimension of EPs from two to one, thus enabling symmetry-protected second-order EPs even in 1D systems \cite{Budich2019,Yoshida2019,Okugawa,Zhou,ips1,ips2}. The generalization of these recent insights and their physical implications to higher-order EPs has largely remained an open problem \cite{NonStandardEP,OzRoNoYa2019,HoHaWiGaElChKh2017,jin-song}. 

\begin{figure} 
	\centerline{\includegraphics[width=0.94\linewidth]{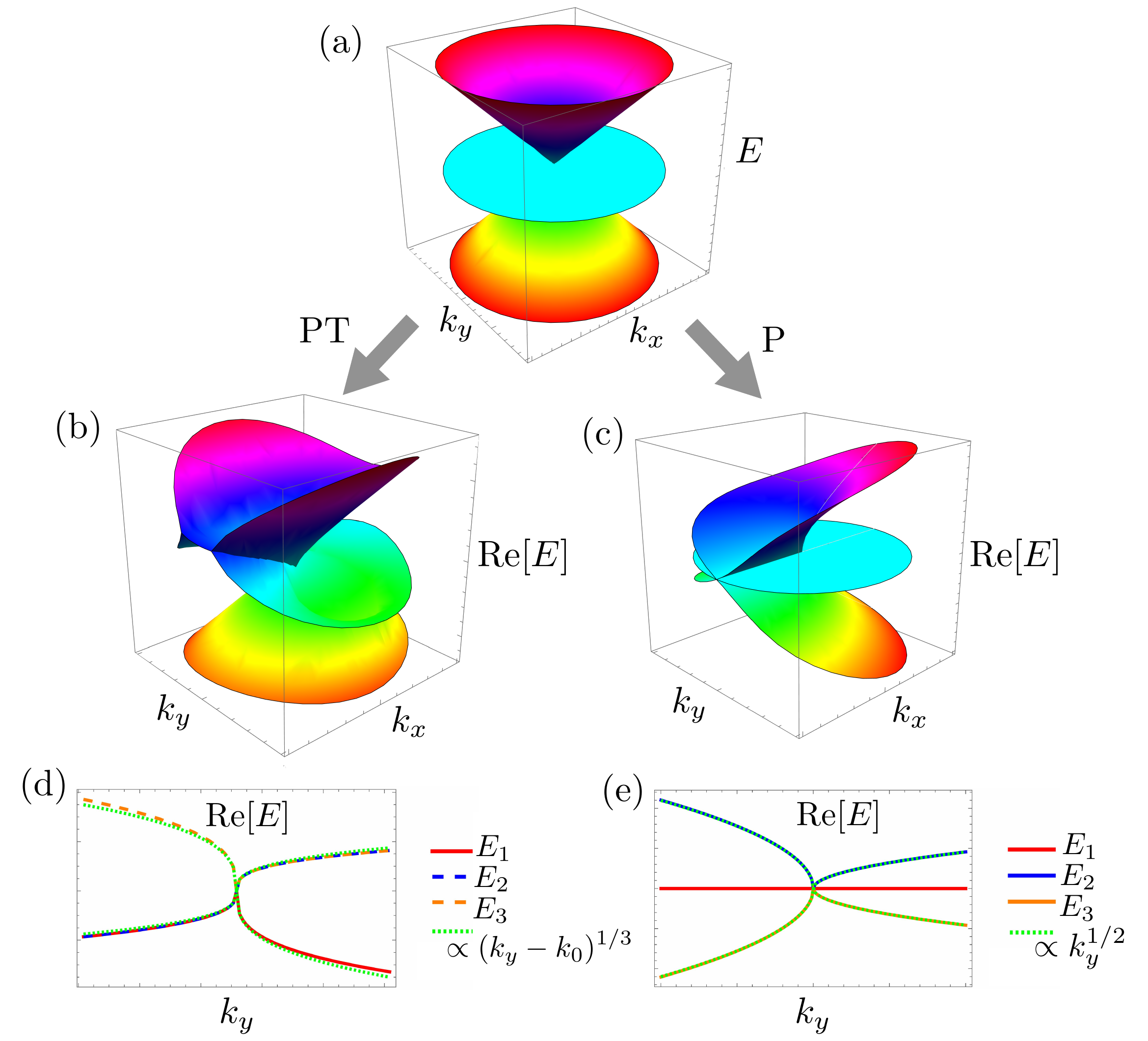}}
	\caption{A Hermitian triple degeneracy (a) ($\epsilon=0$) splits into qualitatively distinct third-order EPs, depending on the symmetry of the (arbitrary small) non-Hermitian perturbation ($\epsilon\neq0$). A PT-symmetric perturbation (Eq. \ref{HPT}) yields EP behaviour featuring generic open surface degeneracies in the (real part of) energy dispersion (b), and a $\sim k^{1/3}$ dispersion away from the 3EPs (d). Distinctly, a P-symmetric perturbation (Eq. \ref{HP}) entails arc degeneracies (c), a perfectly flat band (c,e), and a $\sim k^{1/2}$ dispersion (e) untypical of 3EPs, yet enforced by the symmetry.}\label{fig1main}
\end{figure}

Below, we demonstrate how third-order EPs  (3EPs) naturally occur in 2D systems, stabilized by widely abundant NH symmetries such as parity-time (PT) symmetry. Quite remarkably, this reveals that 3EPs stemming from different symmetries entail qualitatively different phenomenology [see Fig.~\ref{fig1main}]: in addition to the generic $\sim k^{1/3}$ dispersion enforced by PT symmetry, a generalized chiral P symmetry is shown to imply a non-standard dispersion $\sim k^{1/2}$ away from the 3EPs \cite{NonStandardEP} and distinct concomitant arc-degeneracies.

From a mathematical perspective, the findings for PT-symmetric systems are based on the observation that pseudo-Hermiticity constraints (including PT symmetry) reduce the co-dimension of 3EPs by two, i.e. from four to two. This is in stark contrast with the previously studied case of second-order EPs, for which similar symmetries reduce the co-dimension only by one. Again, the P symmetry works differently: instead of having a stable reduction of the co-dimension, the 3EPs show properties reminiscent of fragile and weak topology upon adding more bands and going to higher order. Going beyond the specific case of $n=3$, we generally discuss the abundance of higher-order EPs in symmetric NH systems. Our insights substantially enrich the plethora of NH symmetry protected phases \cite{gong,Budich2019,Yoshida2019,Okugawa,Zhou,lieu2,Kawabata2019,ZhLe2019,
KawabataExceptional}, 
and are of immediate experimental relevance for a wide range of physical settings.


{\it Linearized models.---}
We begin by directly providing simple, linearized, higher-spin Dirac-like NH Hamiltonians 
\begin{equation} 
H_{{\rm PT}}=\begin{pmatrix} 
 -\epsilon  & \mathrm{i} \left (k_x+\epsilon \right ) & -\epsilon \left (1-\mathrm{i} \right )  \\
 \mathrm{i} \left ( -k_x + \epsilon \right ) & 0 & 
 \left (k_y-\epsilon \right ) \left (1+ \mathrm{i} \right ) \\
  \epsilon  \left (1+ \mathrm{i} \right )
  &  \left (k_y+\epsilon \right ) \left (1- \mathrm{i} \right ) & \epsilon  \\
\end{pmatrix},
\label{HPT}
 \end{equation}
and
\begin{equation} H_{{\rm P}}=  \begin{pmatrix} 
0 & \mathrm{i}\,k_x + \mathrm{i} \, \epsilon & 0\\
-\mathrm{i}\,k_x & 0 & \mathrm{i}\,k_y  \\
0 & -\mathrm{i}\,k_y + \epsilon \left(1+\mathrm{i} \right) & 0 
\end{pmatrix} ,
\label{HP}
\end{equation}
describing the 3EPs in 2D with PT symmetry and P symmetry, respectively. At $\epsilon=0$, both models are Hermitian and feature a triple degeneracy, whereby two conical bands touch a flat band at $\mathbf k=0$ [cf. Fig. \ref{fig1main}(a)]. Such degeneracies occur, for instance, in Lieb lattices 
\cite{lieb0,dario,goldman,janik,prb81}, that have been realized in photonic settings \cite{prl1140,prl114} where non-Hermiticities may naturally be introduced (e.g. by staggered losses \cite{NHLieb,Ozawa2019}). Remarkably, for finite $\epsilon$, the degeneracy points immediately split into 3EPs, with a qualitatively different behaviour for the two models, cf. Fig.~\ref{fig1main}. There, we see that while the PT-symmetric model has energies exhibiting a cube-root dispersion, $E\sim k^{1/3}$ [Fig. \ref{fig1main}(b,d)], the P-symmetric model possesses an exactly flat band and two dispersing bands exhibiting a square-root scaling away from the 3EP, viz. $E\sim k^{1/2}$ [Fig.~\ref{fig1main}(c,e)].

\begin{figure*} [t]
	\centerline{\includegraphics[width=0.8\linewidth]{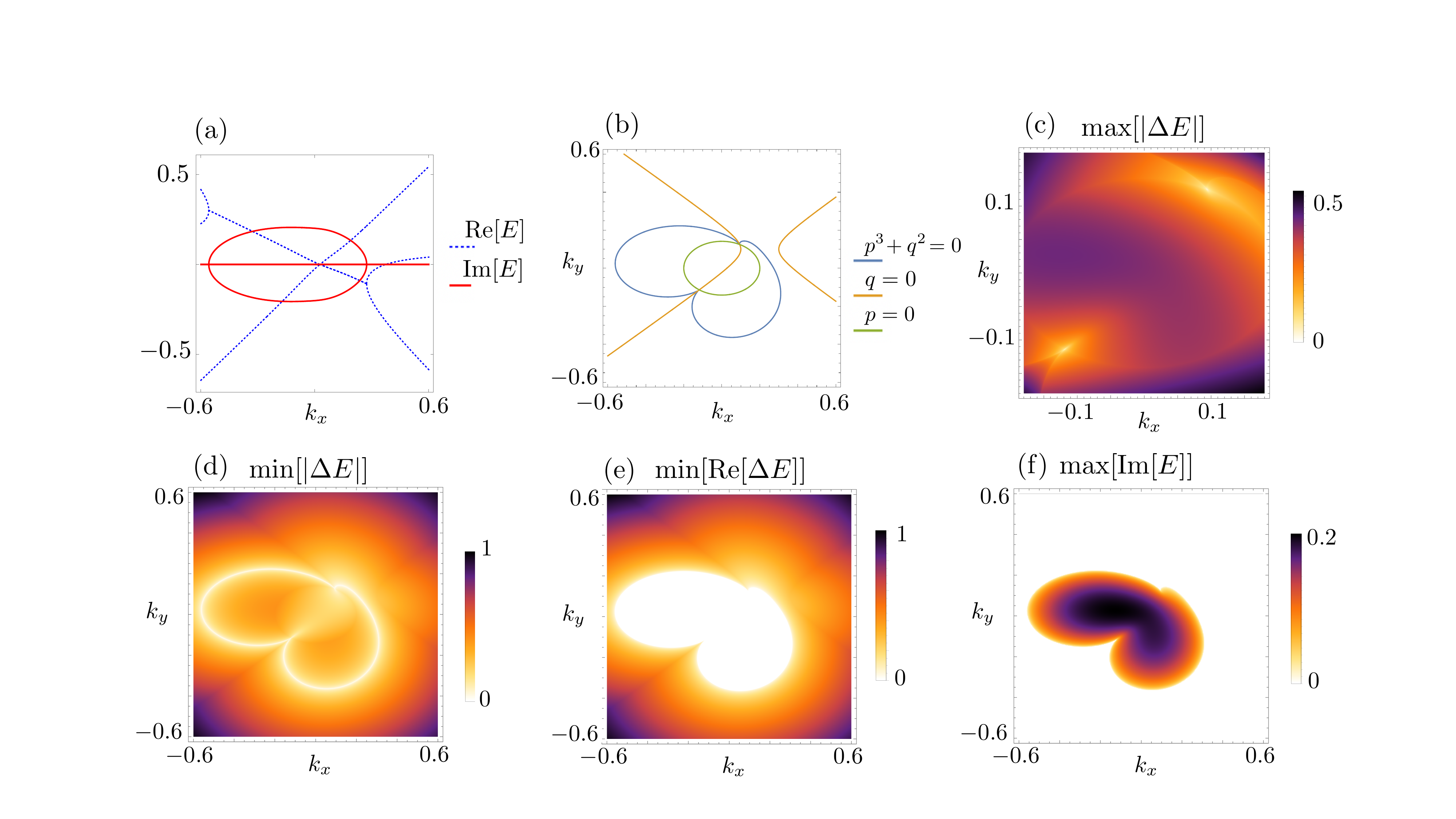}}
	\caption{PT-symmetric model: EPs and generalized arc degeneracies. In all examples, we consider a weak non-Hermiticity quantified by $\epsilon=0.1$ in Eq.~\ref{HPT}. Subfigure (a) shows the general structure of the eigenenergies with an example on the $k_y=0$ line; (b) indicates where the functions $p, q$, and $p^3+q^2$ vanish, which is the key to understanding the topology of the energy bands; (c) highlights the 3EPs by plotting the largest absolute value of the complex gap (note the enlarged scale compared to the other panels); (d) emphasizes the lines of 2EPs by showing the smallest absolute value of the complex gap; (e) and (f) are tailored to highlight the degeneracies in real and imaginary energies, respectively.\label{fig2main}
}	
\end{figure*}

{\it PT symmetry protected 3EPs: Geometry, nodal arcs, and topology.---}
The time-reversal symmetry applied to a non-Hermitian Hamiltonian is also referred to as the K symmetry \cite{bernardleclair,bernardleclair2}. We denote the corresponding unitary matrix as $K$, with $K\, K^* =\pm1$.
We focus on the symmetry classes with $K\,K^* = 1$ and $K\,P^* =  P\,K $, where we denote the P symmetry by a matrix $P$ \cite{bernardleclair,bernardleclair2}. Explicitly, 
$H_{{\rm PT}} (\mathbf k)= P\, K
\left (H_{{\rm PT}}(\mathbf k) \right )^* \left( P\, K \right)^{-1} $, which holds for our Eq.~\ref{HPT} using the representation 
\begin{align}
P= \begin{pmatrix} 
1 & 0 & 0\\
0 & -1 & 0 \\
0 & 0 & 1 
\end{pmatrix},\quad 
K = \begin{pmatrix} 
1 & 0 & 0 \\
0 & 1 & 0 \\
0 & 0 & \mathrm{i} 
 \end{pmatrix}.\label{sym}
\end{align} 
Our PT-symmetric model of 3EPs in Eq.~\ref{HPT} can be fully diagonalized using Cardano's method (details for the general PT-symmetric three-band systems are provided in the Supplemental Material \cite{SOM}). With $\omega = \frac{-1+\sqrt{3} \,\mathrm{i}}{2} $ and
$\alpha_\pm  \equiv \sqrt[3]{ q \pm \sqrt{ p^3+ q^2}}$, we find that
\begin{eqnarray}
E_1&=&\alpha_+ +\alpha_-  \,,\nonumber\\
E_2&=&\omega \, \alpha_+ +  \omega^*  \, \alpha_- \,,\nonumber \\
E_3&=&\omega^* \, \alpha_+ +  \omega  \, \alpha_-\,,
 \label{PTE}
\end{eqnarray}
where $p= \frac{ -k_x^2-2 k_y^2+4 \epsilon ^2 } {3}$
and $q = -\frac{ k_x^2 -2 k_y^2 +4  \left( k_y- k_x \right) \epsilon
+\epsilon ^2 } {2} \epsilon$. It is worth noting that while the expressions appear to be multivalued, the eigenvalues are uniquely defined, hence specifying also the branch cuts. Fig.~\ref{fig2main}(a) displays the generic feature, $\{E_i\}=\{E^*_j \}$, of the spectrum, on a line segment in the Brilluoin zone. In a three-band system, this implies that at each momentum (at least) one energy eigenvalue is always real, while the other two come in complex conjugate pairs with opposite imaginary parts,
viz. ${\rm Im}[E_i]=-{\rm Im}[E_j]$.
The key quantity determining the momentum-space topology is the real-valued quantity $p^3 + q^2$, cf. Fig.~\ref{fig2main}.
It follows that the lines of 2EPs occur on the curve $p^3 + q^2 = 0$ [Fig.~\ref{fig2main}(b)], whereas the 3EPs occur at the special points $p=q=0$ on that curve, which can also be seen as the intersections of the curves $p=0$ and $q=0$ [cf. Fig.~\ref{fig2main}(b) and Fig.~\ref{fig2main}(c)]. When $p^3 + q^2< 0$, we have $E_2=E_3^*$; hence there is a surface on which the real parts of two eigenvalues coincide, ${\rm Re}[E_2]={\rm Re}[E_3]$, forming an open Fermi surface [cf. Fig.~\ref{fig2main}(e)] terminated by the (closed) curves of 2EPs. In the complement of the aforementioned region, i.e. where $p^3 + q^2>0$, all energies are purely real [Fig.~\ref{fig2main}(f)] and distinct [Fig.~\ref{fig2main}(e)]. 

Near the 3EPs at $p=0$ and $q=0$, expanding the expression under the cubic roots in $\alpha_{\pm}$ to linear order, directly leads to the generic dispersion $\sim k^{1/3}$ away from the 3EPs [cf. Fig.~\ref{fig1main}(d)].

The properties described above are generic for PT-symmetric three-band models. Given the ubiquity of PT symmetry, e.g. in photonics \cite{OzRoNoYa2019}, this phenomenology is directly relevant for experiments.

{\it P symmetry protected 3EPs: Lieb lattice model and anomalous dispersion.---}
The P symmetry, acting as $H_{{\rm P}} (\mathbf k)= -P H_{{\rm P}}(\mathbf k) P^{-1}$, is arguably less standard. But importantly, it is naturally obeyed by nearest neighbour hopping models on the Lieb lattice, whose implementations in photonics \cite{prl1140,prl114} provide a strong rationale for studying its extension to the NH realm. We note that, NH perturbations to Lieb lattice models that break P symmetry are also feasible, and that such terms lead to a phenomenology quite different from the one described below \cite{NHLieb}.
Before proceeding to the Lieb lattice model, we consider the basic properties of the P-symmetric three-band models in general. In the presence of a P symmetry, the eigenvalues obey the condition: $\lbrace E_i \rbrace = 
\lbrace - E_j \rbrace $.
 This implies that, for an odd-dimensional matrix, there is always one zero eigenvalue (flat band).
Choosing the unitary P-matrix as in Eq.~\ref{sym},
$H_{{\rm P}}$ is generally restricted to take the form:
\begin{align}
H_{{\rm P}} = \begin{pmatrix} 
0 & b & 0\\
d & 0 & f \\
0 & h & 0 
   \end{pmatrix} ,
\end{align}
with $b,d,f,h\in \mathbb C$. The corresponding eigenvalues are $E=0,\pm\sqrt{b \,d+f \,h}$.
A 3EP is obtained at $ b= -\frac{f\, h}{d}$, with the single normalizable right eigenvector
$\psi_R= (-\frac{f}{d},0,1)^T$ [the left eigenvector is $\psi_L=( -\frac{h}{b},0,1)$]. This shows that although the flat $E=0$ band may seem to play a rather passive role, it, as advertised, enhances the order of the exceptional point.

\begin{figure*} 
	\centerline{\includegraphics[width=0.8\linewidth]{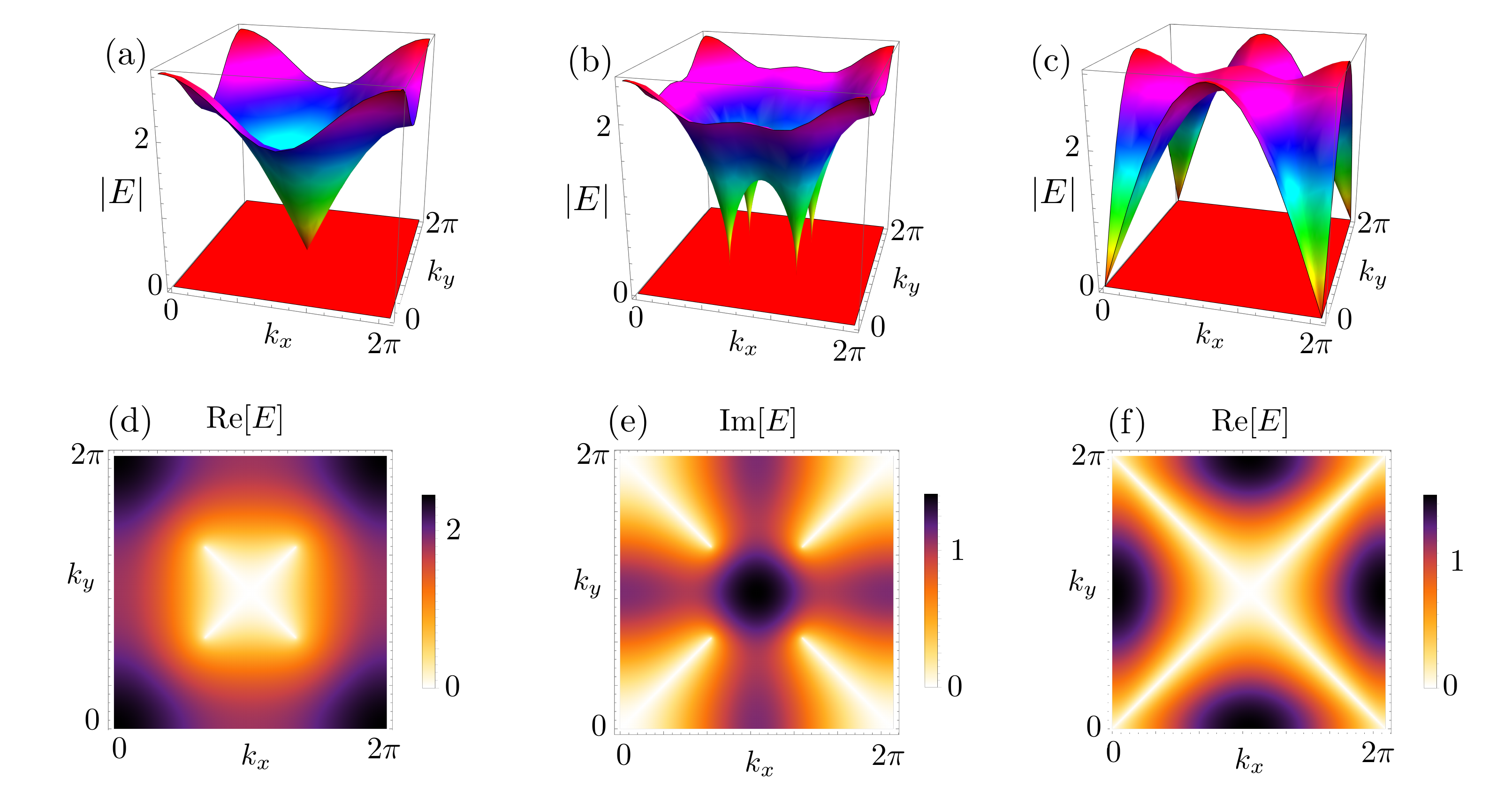}}
	\caption{NH Lieb lattice model with crossed Fermi arcs. A Hermitian degeneracy at $\epsilon=0$ in (a) is split into four 3EPs upon introducing a finite non-Hermiticity, as shown for $\epsilon=1$ in (b), and $\epsilon=2 $ in (c). These 3EPs entail crossed Fermi arc degeneracies (d), and complementary i-Fermi arcs (e). At $\epsilon=2$, the 3EPs recombine (f) before opening up a complex gap at $\epsilon>2$. Furthermore, at $\epsilon=2$, the Fermi arcs (f) override the vanishing i-Fermi arcs, and form closed line degeneracies slicing up the entire Brillouin zone, and persisting at all $\epsilon>2$.}\label{fig3main}
\end{figure*}

To illustrate the generic features of P-symmetric three-band systems we consider an NH Lieb lattice model, 
\begin{equation}
H_{\rm Lieb}\!=\!
  \begin{pmatrix} 
 0 & 1+e^{\mathrm{i} \, k_x}\! -\!\mathrm{i} \,\epsilon & 0 \\
 1+e^{-\mathrm{i}  \,k_x}\! -\!\mathrm{i} \,\epsilon
  & 0 &  1+e^{\mathrm{i}\, k_y}\! +\! \mathrm{i}\, \epsilon \\
 0 & 1+e^{-\mathrm{i} \,k_y}\!+\!\mathrm{i} \,\epsilon & 0 \\
\end{pmatrix},
   \label{hamep3}
\end{equation}
which corresponds to the standard nearest neighbour model at $\epsilon=0$ \cite{lieb0,dario,goldman,prb81,prl1140,prl114} [cf. Fig. \ref{fig1main}(a) and Fig. \ref{fig3main}(a)]. The corresponding eigenvalues are $0$ and $ \pm E$, where
\begin{equation}
E\! =\!
\sqrt{2\left( 2+\!\cos k_x\! +\!\cos k_y-\! \mathrm{i} \, \epsilon  \cos k_x
\! +\! \mathrm{i} \, \epsilon  \cos k_y\! 
-\epsilon ^2\!\right)  }\,.
\end{equation}
The 3EPs are located 
at $\left \lbrace   k_x=\pm \cos ^{-1}\left(-1 + \frac{\epsilon^2}{2} \right), 
k_y = \pm \cos ^{-1}\left( -1+ \frac{\epsilon^2}{2} \right)\right \rbrace $. These develop symmetrically in the BZ with increasing $\epsilon$, originating from the Hermitian ($\epsilon=0$) degeneracy at $\left \lbrace   k_x=\pi, 
k_y =\pi\right \rbrace $, as shown for $\epsilon=0,1$, and $2$ in Fig.~\ref{fig3main}(a), (b), and (c), respectively. At $\epsilon=2$, the 3EPs again recombine to a linear band-touching, before developing a complex gap for $\epsilon>2$. Note also that for $0<\epsilon<2$, the dispersion away from the 3EPs has a square-root behaviour, $\sim k^{1/2}$. Such a behaviour is typically associated with 2EPs, but is known as a mathematical possibility \cite{NonStandardEP} for higher-order EPs. Here it is enforced for 3EPs by the P symmetry.

Accompanying the 3EPs in our model, there are arc-like degeneracies in the real and imaginary parts of the spectrum (see Fig.~\ref{fig3main}(d) and (e), respectively), again reminiscent of the ``Fermi arcs" accompanying the more conventional 2EPs occurring in 2D models without symmetry \cite{koziifu,NHarc}. In fact, the arcs in the P-symmetric model are arguably more generally and directly relevant to experiments as they are constrained to occur at Re$[E]=0$ while in two-band models the degeneracy lines (arcs) themselves they are in general dispersive. This should greatly facilitate the observation of arcs in light scattering experiments similar to those of Ref. \onlinecite{NHarc}, here possibly being much more extended than the beautiful but minuscule arcs observed as a consequence of 2EPs. 

Moreover, the structure of these arcs is highly unusual: they connect opposite corners, and cross at $k_x=k_y=\pi$ [cf. Fig.~\ref{fig3main}(d)]. The complementary ``i-Fermi arcs'' \cite{review} connect the same points, exploiting the periodicity of the Brillouin zone [cf. Fig. \ref{fig3main}(e)]. As $\epsilon\rightarrow 2$, the i-Fermi arcs give way to the (real) Fermi arcs, that join to form closed degeneracy
lines slicing through the entire Brillouin zone [cf. Fig.~\ref{fig3main}(f)]. Intriguingly, these line degeneracies in Re$[E]$ remain as vestiges of the Fermi arcs at any $\epsilon >2$.

{\it Higher order and number of bands: Coexistence, strong, weak and fragile aspects.---} For EPs of dimension lower than the number of bands in the system, the PT and P symmetries have radically different implications. For PT symmetry, we have a ``stable'' situation in the sense that the sufficient number of tuning parameters remains the same, regardless of the number of additional bands. A manifestation of this is the emergence of lines of 2EPs in our PT-symmetric three-band model, studied above (cf. Ref. \onlinecite{Budich2019}). The EPs in the P symmetry case are, in contrast, fragile in the sense that they can even be forbidden when additional bands are added. For three bands, we saw that no 2EPs are possible. Similarly, no 3EPs occur in four-band models \cite{SOM}. In analogy with weak topological insulators, there is, however, an odd-even effect: the 3EPs may again be realized by tuning just two parameters in P-symmetric five-band models.

Allowing more tuning parameters makes the situation even richer. For instance, in analogy with the two-band case without symmetries \cite{CaStBuBe2019,chinghuaknot,StRoArBuBe2019,yangknot}, knotted lines of 3EPs should occur as stable features in both PT- and P-symmetric NH models in 3D. In the PT-symmetric case, these will generically coexist with surfaces of 2EPs. This line of reasoning opens up the possibility for a rich variety of coexisting EP motifs of different orders and dimensionalities, generalizing the situation to coexisting 2EP lines and 3EPs (as seen in our explicit PT-symmetric 2D example).  

{\it Discussion.---} In this Letter, we have shown that symmetries not only make higher-order exceptional degeneracies much more abundant (and thus physically more relevant), but also results in the fact that different underlying symmetries can lead to exceptional degeneracies associated with strikingly different phenomenology. This sets the stage for several directions of further research. On a theoretical level, realistic example models in various dimensions and at various orders deserve detailed investigation, including their potentially anomalous open-boundary physics (which is known to be a rather generic property of topological NH systems \cite{lee, BBC,yaowang}).

The generality of the argumentation, and the ubiquity of non-Hermiticity in a wide range of dissipative metamaterials also make the experimental exploration of higher-order EPs a very promising avenue. In this context, we note that very soon after their theoretical prediction, both symmetry protected 2EP phases and even seemingly quite complicated models of knotted 2EP-lines, were tailor-made in optics experiments \cite{knotexp}. There is no obvious reason preventing the symmetry protected 3EPs discovered here, and even knotted 3D generalizations thereof, to be realized and observed using similar setups.


\acknowledgments
{\it Acknowledgments.---}
We are grateful to Jan Budich for insightful discussions. I.M. and E.J.B. are supported by the Swedish Research Council (VR) and the Wallenberg Academy Fellows program of the Knut and Alice Wallenberg Foundation.

{\it Note added.---} Upon submission of this manuscript, a related work appeared \cite{3Tsuneya}. While their general analysis is highly relevant for our PT symmetry analysis, they did not consider P symmetry, and thus did not identify the emergence of the non-standard higher-order EPs (such as the 3EPs with Fermi arcs and a square-root dispersion).



\newpage
\begingroup
\onecolumngrid
\appendix
\section{Supplemental Material for ``Symmetry and Higher-Order Exceptional Points''}

In this supplemental material, we provide the details of our derivations and arguments presented in the main text.  

\subsection{Conditions required to get an $N^{th}$ order degeneracy for a generic $N\times N $ complex matrix}

Any complex square matrix is triangularizable (Schur’s lemma). The eigenvalues are the diagonal entries of this upper or lower triangular form.
Hence, a generic $N\times N $ non-Hermitian complex matrix can be brought into the form $ \begin{pmatrix} 
E_1 & 0 & \dots & 0  \\
\kappa_{21} & E_2 & \dots &  0\\
\vdots & \vdots & \ddots & \vdots \\
\kappa_{N1} & \kappa_{N2} & \dots & E_N
   \end{pmatrix}$,
where all the entries  are complex in general.
A special case is the matrix
$\begin{pmatrix} 
E & 0 & \dots & 0  \\
\kappa_{21} & E & \dots &  0\\
\vdots & \vdots & \ddots & \vdots \\
\kappa_{N1} & \kappa_{N2} & \dots & E
   \end{pmatrix}$, 
which has an $n^{\rm th}$-order Exceptional Point ($n$EP) with $n=N$. It is defective, having a single eigenvalue $E$ with algebraic multiplicity $N$, i.e. the $N$ eigenvalues coalesce.  
Clearly, we need $(N -1 )$ complex constraint equations to get the $N^{th}$-order EP.

As a simple example, let us consider a  two-level non-Hermitian system with a Hamiltonian of the form
$
 \begin{pmatrix} 
a & c \\
d & b 
   \end{pmatrix}$, with complex parameters $a,b,c,d$.
Since the eigenvalues are given by $\frac{ a+b \pm \sqrt{(a-b)^2+4 \, c\, d}} {2}$,
a $2$EP can be obtained under the condition $(a-b)^2+4 \,c \,d=0$. 
Hence, we need $2$ real constraints to get an EP from
a generic $2\times 2 $ complex matrix.

\subsection{PT-symmetric $3\times 3$ non-Hermitian matrix}

We refer to the time-reversal symmetry applied to a non-Hermitian Hamiltonian as the K symmetry, and denote the corresponding unitary matrix as the matrix $K$, where $ K\, K^* =\pm 1$.
We focus on the symmetry classes with $K\,K^* = 1$  and $ K\,P^* =  P\, K $, where we denote the P symmetry by the matrix $P$.
For the $3\times 3$ case, we can choose the representations for $P$ and $K$ as:
\begin{align}
P = \begin{pmatrix} 
1 & 0 & 0\\
0 & -1 & 0 \\
0 & 0 & 1 
\end{pmatrix},\text{ and } 
K = \begin{pmatrix} 
1 & 0 & 0\\
0 & 1 & 0 \\
0 & 0 & \mathrm{i} 
 \end{pmatrix},
\end{align} 
respectively.
Under the combined PT symmetry, a three-band complex Hamiltonian is then restricted to take the form:
\begin{align}
H_{\text{PT}}  = \begin{pmatrix} 
 real_1 & \mathrm{i} \,real_2 & real_3 \left(1 - \mathrm{i} \right ) \\
\mathrm{i}\, real_4 & real_5 & real_6  \left(1 + \mathrm{i} \right ) \\
 real_7 \left(1+\mathrm{i} \right ) & real_8 \left(1-\mathrm{i} \right ) & real_9 
   \end{pmatrix},
\end{align}
which satisfies the condition $ H_{\text{PT}} = \left( P\, K \right)
 H_{\text{PT}}^* \left( P\, K \right)^{-1} $.
Here, the parameters $real_1, real_2, real_3, real_4, real_5, real_6, real_7, real_8,$
and $real_9$ are all real.
The eigenvalues of $H_{\text{PT}}$ can be found using Cardano's method, and are given by:
\begin{align}
\alpha_+ +\alpha_- + \beta \,,\quad
\omega \, \alpha_+ +  \omega^*  \, \alpha_- + \beta \,,\quad
\omega^* \, \alpha_+ +  \omega  \, \alpha_- + \beta\,,
\end{align}
where
\begin{align}
& 
\alpha_\pm  = \sqrt[3]{ \tilde q \pm \sqrt{\tilde p^3+ \tilde q^2}}\,,
\quad 
\omega =\frac{-1+\sqrt{3} \,\mathrm{i}}{2} \,,\quad
 \beta =  \frac{real_1 +real_5+real_9} {3} \,,\nn
&   p =\frac{
-real_1^2 + 3 \,real_2 \,real_4 - real_5^2 - 6\, real_3 \,real_7 
- 6 \,real_6 \,real_8 + real_5\, real_9 - real_9^2 + real_1 \left (real_5 + real_9 \right)}
{9} \,,\nn
&   q= \frac{ \left (real_1 + real_5 + real_9 \right )^3}  {27}
-real_3\, real_5 \,real_7 
- real_2\, real_6\, real_7 
+ real_3 \,real_4\, real_8 - real_1\, real_6\, real_8 \nn
& \qquad 
 - \frac{ (real_1 + real_5 + real_9) 
 \left[ real_2\, real_4 - 2\, real_3 \,real_7 
 - 2\, real_6\, real_8 + real_5\, real_9 + 
 real_1 \left (real_5 + real_9 \right ) \right ]} {6}
\nn & \qquad
+ \frac{real_1 \,real_5\, real_9+ real_2\, real_4\, real_9} {2}     \,.
\end{align}
For this case, we need 2 real constraints, namely $p=0$ and $q=0$, to get 3EPs.
Around a 3EP, the eigenvalues show a leading dependence of
$\sim \left(k-k_0 \right)^{\frac{1}{3}}$.

\subsection{PT-symmetric $4\times 4$ non-Hermitian matrix}

If we choose the representations of the matrices $P$ and $K$ as
\begin{align}
\label{eqP4band}
P=\begin{pmatrix}
 1 & 0 & 0 & 0 \\
 0 & -1 & 0 & 0 \\
 0 & 0 & 1 & 0 \\
 0 & 0 & 0 & -1
\end{pmatrix}, \text{ and }
K=\begin{pmatrix}
 0 & 0 & 1 & 0 \\
 0 & 0 & 0 & 1 \\
 1 & 0 & 0 & 0 \\
 0 & 1 & 0 & 0
\end{pmatrix},
\end{align}
respectively, we get the symmetry classes obeying $ K\,K^* = 1$ and $K\,P^* =  P\,K $.
Under the combined PT symmetry,  a four-band matrix is then restricted to take the form:
\begin{align}
H_{\text{PT}} = \begin{pmatrix} 
 real_1- \mathrm{i} \, imag_1 
 & -(real_2-\mathrm{i} \, imag_2) 
 & real_3-\mathrm{i} \, imag_3 & real_4+ \mathrm{i} \, imag_4 \\
 -(real_5-\mathrm{i} \, imag_5) & real_6-\mathrm{i} \, imag_6 & -(real_7-\mathrm{i} \, imag_7) & real_8-\mathrm{i} \, imag_8 \\
 real_3+ \mathrm{i} \, imag_3 & -(real_4-\mathrm{i} \, imag_4) & real_1+\mathrm{i} \, imag_1 
 & real_2+ \mathrm{i} \, imag_2 \\
 real_7+ \mathrm{i} \, imag_7 & real_8+ \mathrm{i} \, imag_8 & real_5
 +\mathrm{i} \, imag_5 & real_6+ \mathrm{i} \, imag_6 \\ 
 \end{pmatrix},
\end{align}
where the parameters $real_1, real_2, real_3, real_4, real_5, real_6, real_7, real_8,
imag_1, imag_2, imag_3, imag_4, imag_5 , imag_6, imag_7$, and $ imag_8$
are all real.
Let us consider $-\det\left[ H_{\text{PT}} - \lambda \,  I_{4\times 4} \right]$,
where $\beta$ is the coefficient of $\lambda^3$, $ \gamma$ is the coefficient of $\lambda^2$, $\delta $ is the coefficient of $\lambda$, and $ \varepsilon $ is the $\lambda$-independent term. 
The associated depressed quartic is given by $y^4+ \tilde{p}\, y^2 +
\tilde q \, y+ \tilde r = 0, $ where 
\begin{align}
\tilde p =  \gamma -\frac{3\,\beta^2}{8}\,,\quad
\tilde q = \delta +\frac{ \beta^3-4 \,\beta \,\gamma } {8}\,,\quad
\tilde r  = 
\frac{-3\,\beta ^4 +256 \,\varepsilon -64 \,\beta\,\delta 
+16 \, \gamma \,\beta ^2}
{256} \,.
\end{align}
The four eigenvalues are given by the four roots of the associated depressed quartic in $y$ shifted by $-\beta /4$.
For all the roots to coincide, we then need $\tilde p =\tilde q =\tilde r= 0$.
Evaluating the expressions for these quantities for $H_{\text{PT}}$ show that $\tilde p, \tilde q$, and $ \tilde r$ are real.
This proves that we need to satisfy 3 real constraints in order to get a 4EP.
 
\subsection{P-symmetric $4\times 4$ non-Hermitian matrix}

Using the representation in Eq.~\ref{eqP4band}, a P-symmetric $4\times 4$ non-Hermitian matrix is restricted to the form:
\begin{align}
H_{\text{P}} = \begin{pmatrix} 
0 & b & 0 & j \\
d & 0& f & 0 \\
0 & h & 0 & l \\
m & 0 & o & 0 \\
\end{pmatrix},
\end{align}
where all parameters are in general complex.
The eigenvalues are then given by:
\begin{align}
 \pm 
\sqrt{ \frac{
\left( b \,d+ f \,h+  j \,m+l \,o \right)
\pm \sqrt{ 
\left (b \,d+ f\, h + j \, m+ l \, o \right )^2
- 4 \left  ( h\, j-b\, l \right ) 
\left ( f \,m-d \, o \right )}
} {2} } \,.
\end{align}
This proves that we need 2 complex conditions to get a 4EP.


\end{document}